\documentclass[prl,twocolumn,amsmath,amssymb,superscriptaddress]{revtex4-1}
\usepackage{graphicx}
\usepackage{dcolumn}
\usepackage{bm}
\usepackage{amsmath}
\usepackage{amsfonts}
\usepackage{amssymb}
\usepackage{natbib}

\begin{document}

\title{Experimental Optimum  Maximum-Confidence Discrimination and\\
Optimum Unambiguous Discrimination of Two Mixed Single-Photon States}

\author{Gesine A. Steudle}
\email{steudle@physik.hu-berlin.de}
\affiliation{Humboldt-Universit\"{a}t zu Berlin, AG Nanooptik, Newtonstr. 15, 12489 Berlin, Germany}
\author{Sebastian Knauer}
\affiliation{Humboldt-Universit\"{a}t zu Berlin, AG Nanooptik, Newtonstr. 15, 12489 Berlin, Germany}
\author{Ulrike Herzog}
\affiliation{Humboldt-Universit\"{a}t zu Berlin, AG Nanooptik, Newtonstr. 15, 12489 Berlin, Germany}
\author{Erik Stock}
\affiliation{Technische Universit\"{a}t Berlin, Institut f\"{u}r Festk\"{o}rperphysik, Hardenbergstr. 36, 10623 Berlin, Germany}
\author{Vladimir A. Haisler}
\affiliation{Institute of Semiconductor Physics, Lavrenteva avenue 13, Novosibirsk 630090, Russia}
\author{Dieter Bimberg}
\affiliation{Technische Universit\"{a}t Berlin, Institut f\"{u}r Festk\"{o}rperphysik, Hardenbergstr. 36, 10623 Berlin, Germany}
\author{Oliver Benson}
\affiliation{Humboldt-Universit\"{a}t zu Berlin, AG Nanooptik, Newtonstr. 15, 12489 Berlin, Germany}
\homepage{http://www.physik.hu-berlin.de/nano}

\date{\today}

\begin{abstract}
We present an experimental implementation of optimum measurements for quantum state discrimination. Optimum maximum-confidence discrimination and optimum unambiguous discrimination of two mixed single-photon polarization states were performed. For the latter the states of rank two in a four-dimensional Hilbert space are prepared using both path and polarization encoding. Linear optics and single photons from a true single-photon source based on a semiconductor quantum dot are utilized. 
\end{abstract}


\maketitle

The discrimination between quantum states in a given set of states is a fundamental challenge in quantum communication and quantum cryptography \cite{Barnett2009}. The straightforward way to obtain information about a quantum state is a projective measurement on an orthogonal basis (von Neumann measurement). For nonorthogonal states, where perfect discrimination is impossible, different optimum measurement strategies have been theoretically derived and experimentally demonstrated using linear optics \cite{Barnett2009}. When two input states $\rho_1$ and $\rho_2$ with respective a priori probabilities $\eta_1$ and $\eta_2$ are to be discriminated without inconclusive results, the minimum overall error probability, $P_E$, is achieved by a projective measurement on a certain orthogonal basis, yielding the Helstrom bound $P_E=\frac{1}{2}(1-\mathrm{Tr}|\eta_1 \rho_1-\eta_2 \rho_2|)$ \cite{Helstrom1976}.

At the expense of admitting inconclusive results where the measurement fails to give a definite answer, two nonorthogonal density operators $\rho_{1}$ and $\rho_{2}$ can be probabilistically discriminated without errors, that is unambiguously, provided that their supports \cite{Note1} are different. The optimum measurement for unambiguous discrimination is the one which yields the minimum probability of inconclusive results, $Q_{opt}$. For two nonorthogonal pure states $|\psi_{1}\rangle$ and $|\psi_{2}\rangle$ with equal prior probabilities $\eta_1=\eta_2=0.5$ this minimum is given by $Q_{opt}=|\langle\psi_{1}\vert\psi_{2}\rangle|$ \cite{Ivanovic1987,Dieks1988,Peres1988}. It is achieved by a
generalized measurement that has been implemented experimentally \cite{Huttner1996, Clarke2001a}. The optimum measurement for unambiguously discriminating between a pure state and a mixed state was also derived \cite{Bergou2003} and experimentally realized \cite{Mohseni2004}. For distinguishing two mixed states, complete solutions for the optimum measurement have been obtained for special cases \cite{Rudolph2003,Herzog2005, Raynal2005, Bergou2006, Herzog2007,Kleinmann2010} and recently a classification of the measurements yielding optimum unambiguous discrimination between two completely arbitrary mixed states has been given \cite{Kleinmann2010}.

For the case when unambiguous discrimination is impossible, the strategy of maximum confidence discrimination has been introduced \cite{Croke2006} and experimentally realized for three linearly dependent pure states \cite{Mosley2006}. The confidence $C_{j}$ in the conclusive measurement outcome $j$ is defined as the conditional probability $P(\rho_{j}\vert j)=P(\rho_{j},j)/P(j)$ that the state $\rho_{j}$ was indeed prepared, given that the outcome $j$ is detected \cite{Croke2006}. In other words, $C_{j}$ is the ratio of the number of instances when the outcome $j$ is correct and the total number of instances when the outcome $j$ is detected. If $C_{j}=1$ for all $j$ the measurement is unambiguous. The optimum maximum-confidence measurement is the one that yields the minimum overall probability of inconclusive results, $Q_{opt}$, while the confidence $C_{j}$ is maximal for all $j$.  For discriminating two mixed states this optimum measurement has been theoretically studied \cite{Herzog2009}.

In this paper we report on the experimental realization of optimum unambiguous discrimination between two mixed states, based on a previous proposal \cite{Herzog2010}. The states we discriminate belong to a special class of similar states, $\rho_{2}=U\rho_{1}U^{\dagger}$, where the unitary operator $U$ can be decomposed into rotations in two-dimensional subspaces. States of this kind have been proposed for secure quantum cryptography based on two mixed states \cite{Koashi1996} and the optimum measurement for unambiguously discriminating them  has been theoretically derived \cite{Herzog2007}. We utilize photons from a quantum-dot based true single-photon source to produce the
mixed states, represented by the photon polarization and its spatial paths. So far state discrimination experiments \cite{Huttner1996, Barnett1997, Clarke2001a, Clarke2001b, Mizuno2001, Mohseni2004, Mosley2006} were mostly based on photons from a laser beam, except for two very recent ones using photons from parametric down conversion for extending the strategy of minimum-error discrimination to the availability of multiple copies \cite{Higgins2009} and to entangled states \cite{Lu2010}. Applying a part of our experimental setup, we also implement a proposed measurement \cite{Herzog2010} for discriminating with maximum confidence and minimum probability of inconclusive results between two mixed single-photon polarization states for photons in the same path.

Let us first outline the basic ideas. In our experiment state preparation for unambiguous discrimination starts from a general mixed single-photon state
\begin{equation}
\rho_{0}\! =\! r_{11}\vert H\rangle_{1}\langle H\vert_{1}+r_{22}\vert V\rangle_{2}\langle V\vert_{2} + \left(r_{12}\vert H\rangle_{1}\langle V\vert_{2} + {\rm H. a.}\right) \label{rho0}
\end{equation}
where $|H\rangle_i$ and $|V\rangle_i$ refer to horizontal and vertical polarization, respectively, and the indices $i=1,2$ correspond to the two output ports of a polarizing beam splitter. The state $\rho_0$ is transformed with equal probability $\eta_1=\eta_2=0.5$ either into the state $\rho_1$ or into the state $\rho_2$ where
\begin{equation}
\rho_{1}=U^{(+)}\rho_{0}U^{(+)\dagger}, \quad \rho_{2}=U^{(-)}\rho_{0}U^{(-)\dagger}.
\label{rho12}
\end{equation}
Here  the unitary transformations $U^{(+)}$ and $U^{(-)}$ are composed of rotations by an angle $\alpha$ in two mutually orthogonal two dimensional subspaces, spanned by the states $|H\rangle_1$ and $|V\rangle_1$, on the one hand, and $|H\rangle_2$ and $|V\rangle_2$, on the other hand. More precisely, we have $U^{(\pm)}= U_1^{(\pm)} \otimes U_2^{(\pm)}$ with 
\begin{eqnarray}
U_1^{(\pm)}\vert H\rangle_{1}&=& \cos{\alpha}\vert H\rangle_{1}\pm \sin{\alpha}\vert V\rangle_{1}\equiv\vert r_{\pm}\rangle\\
U_2^{(\pm)}\vert V\rangle_{2} &=& \cos{\alpha}\vert H\rangle_{2}\pm\sin{\alpha}\vert V\rangle_{2}\equiv\vert s_{\pm}\rangle \label{U}
\end{eqnarray}
and $0\le\alpha\le\frac{\pi}{4}$. Clearly, the states $\rho_1$ and $\rho_2$ are two mixed states of rank 2 the supports of which jointly span a four dimensional Hilbert space. It has been shown \cite{Herzog2007,Herzog2010} that for these particular mixed states unambiguous discrimination with minimum failure probability is achieved in a generalized measurement which performs an optimum unambiguous discrimination between the pairs of pure states $\vert r_{\pm}\rangle$ on the one hand and $\vert s_{\pm}\rangle$ on the other hand, yielding the minimum failure probability $Q_{opt} = \cos 2\alpha$ which does not depend on $\rho_0$.

Before entering the polarizing beam splitter leading to the state (\ref{rho0}), the photon is generated in a general mixed single-photon polarization state which can be always written as
$\rho=p\vert\psi\rangle\langle\psi\vert+(1-p)\frac{I_{2}}{2}$. Here $p$ with $0<p\le 1$ is the degree of polarization, $|\psi\rangle$ denotes some pure polarization state, and $I_{2}$ is the identity operator in the two-dimensional Hilbert space. Different mixed states of this kind have identical supports and cannot be unambiguously discriminated. The maximum confidences $C_+$ and $C_-$ for discriminating two states $\rho_+$ and $\rho_- $ defined by
\begin{equation}
\rho_{\pm}=p \vert \psi_{\pm} \rangle \langle \psi_{\pm} \vert+(1-p)\frac{I_{2}}{2},\quad|\psi_{\pm} \rangle=\cos{\beta}\vert H\rangle\pm \sin{\beta}\vert V\rangle \label{part}
\end{equation}
and having equal prior probabilities are \cite{Herzog2009}
\begin{equation}
\label{cmax} C_+ = C_- =\frac{1}{2}+\frac{p\,\sin{2\beta}}{2\sqrt{1-p2\cos2{2\beta}}}.
\end{equation}
These confidences are achieved in a generalized measurement, at the expense of a minimum probability of inconclusive results, given by $Q_{opt}=p \cos{2\beta}$ \cite{Herzog2009,Herzog2010}. By contrast to this, the confidences which are obtained  without inconclusive results when the projective minimum-error measurement is applied are $C_+ ^E =C_- ^E = \frac{1}{2}(1+p\,\sin 2\beta)$ \cite{Herzog2009}.

In our experiment we present an experimental realization of an unambiguous state discrimination measurement for two mixed single-photon states, generated from a single-photon source. The source consists of an electrically pumped single InGaAs/GaAs quantum dot (QD), embedded in a microcavity. The device has a high single photon emission rate and is capable for operating at repetition rates of up to 1 GHz \cite{Lochmann2009}. In this experiment we used cw-excitation.

\begin{figure}
\includegraphics[width=8.5cm]{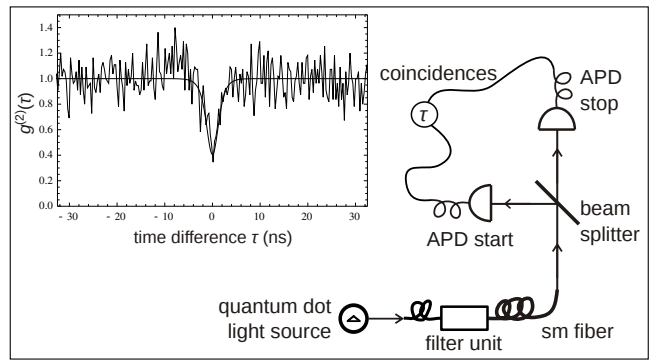}
\caption{HBT setup, the inset shows the measured second order correlation function and a fit for the data. The remaining counts at $\tau=0$ are due to the time resolution of the HBT-setup of 800 ps, which is of the same order of magnitude as the characteristic time-scale of the QD source.}
\label{HBT}
\end{figure}

The light from the QD sample is coupled into a single mode optical fiber. For spectral filtering of a single excitonic transition in the QD, it passes a fiber-coupled filter unit which contains a longpass and a tunable bandpass of a width of 1 nm (Fig. \ref{HBT}). To determine the photon statistics of the light source, the output of the filter unit was coupled into a  Hanbury-Brown and Twiss (HBT)-Setup (Fig. \ref{HBT}). The inset of Fig. \ref{HBT} shows the coincidences against the time difference $\tau$ in detection events in either of the two avalanched photodiodes APD start and APD stop. Clearly an antibunching dip with $g^{(2)}(0)=0.35$, i.e. nonclassical photon statistics, is observed.

\begin{figure}
\includegraphics[width=7cm]{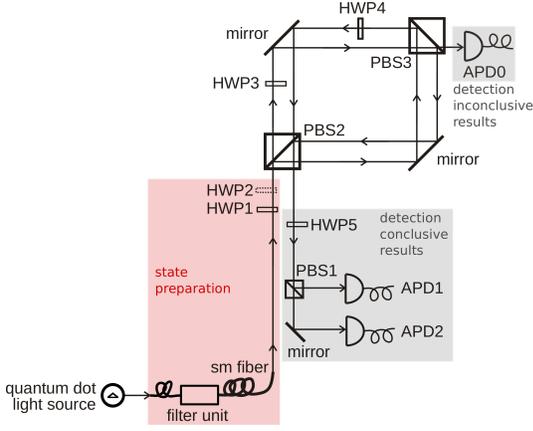}
\caption{Setup for the optimum maximum confidence discrimination measurement for two equally probable single-photon polarization states $\rho_+$ and $\rho_-$ of the same purity, as given in (\ref{part}). HWP denotes half wave plates, PBS polarizing beam splitters, and APDs are single-photon detectors (avalanche photodiodes).}
\label{OMC-setup}
\end{figure}

We experimentally verified that the light from the QD has no preferential polarization. However, after the fiber-coupled filter unit it is in a partially linearly polarized state $\rho=p\vert\psi\rangle\langle\psi\vert+(1-p)\frac{I_{2}}{2}$ where $p=0.54$ was measured.  In order to  test a part of our experimental setup (Fig. \ref{OMC-setup}), we first applied the measurement scheme that was proposed for discriminating two partially polarized single-photon states with the same degree of polarization optimally and with maximum confidence \cite{Herzog2010}. At the half-wave plates HWP1 (HWP1 and HWP2)  the state $\rho_{+}$ ($\rho_{-}$) as given by (\ref{part}) is prepared with equal prior probability. At HWP3 the polarization is turned by
\begin{equation}
\theta_{3}=\arccos{\sqrt{\frac{2 p \cos{2\beta}}{1+p\cos{2\beta}}}},
\end{equation}
at HWP4 by $\theta_{4}=\frac{\pi}{2}$, and at HWP5 by $\theta_{5}=\frac{\pi}{4}$ as proposed in \cite{Herzog2010}. The results are collected at three avalanche photodiodes (APDs), the inconclusive results at APD0 and the conclusive ones at APD1, indicating the presence of $\rho_+$, and at APD2 for $\rho_-$, respectively.

\begin{figure}
\includegraphics[width=6cm]{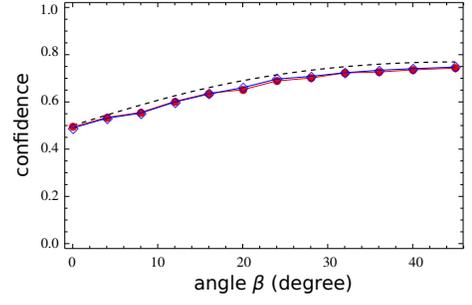}
\caption{Measured confidences $C_+$ and $C_-$ against the half angle of separation $\beta$ for the cases that $\rho_+$ (circles) and $\rho_-$ (diamonds) were prepared. The dashed line shows the theoretical curve.}
\label{OMC}
\end{figure}

Fig. \ref{OMC} shows the measurement results for the confidence. The theoretical curve is given by (\ref{cmax}). Our experimental results follow the expected behaviour. However, for $p=0.54$ the theoretical confidences $C_{\pm}$ for the implemented optimum maximum confidence measurement and the confidences $C_{\pm}^E$ arising from the projective minimum-error measurement differ in less than 0.02 for all angles $\beta$. This difference is too small to be resolved by the experimental data. Yet the experiment confirms stable operation of the complex optical setup over days which is required for the small single-photon signal, a total count rate of around 2000 $\mathrm{s}^{-1}$ was measured when using the true single-photon source.

\begin{figure}
\includegraphics[width=7cm]{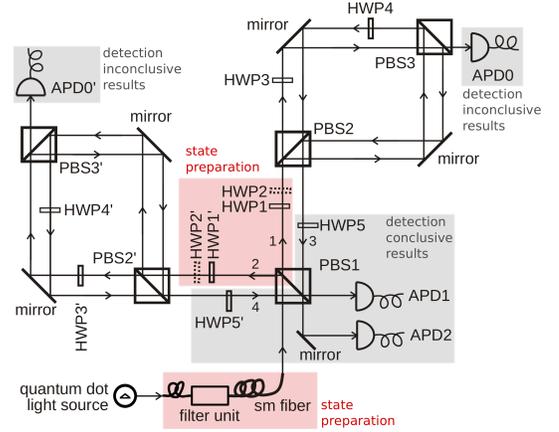}
\caption{Setup for the unambigiuous state discrimination measurement for two mixed single-photon states $\rho_1$ and $\rho_2$, as given in (\ref{rho12}).}
\label{USD-setup}
\end{figure}

For the realization of unambiguous state discrimination with two mixed single-photon states the setup shown in Fig. \ref{OMC-setup} has to be extended by adding a second path \cite{Herzog2010}. This setup is shown in Fig. \ref{USD-setup}. The incoming light passes a first polarizing beam splitter (PBS1). The photon state after PBS1 is given by (\ref{rho0}). The transformation $U^{(+)}$ or $U^{(-)}$ is done at half-wave plates HWP1, HWP1$'$, HWP2, HWP2$'$. At HWP1 and HWP1$'$ the polarization is turned by $\alpha$ and $-(90°-\alpha$), respectively. Thus the state $\rho_{1}$ is prepared according to (\ref{rho12}). To prepare $\rho_{2}$ HWP2 and HWP2$'$ are inserted additionally. After preparation of either $\rho_{1}$ or $\rho_{2}$, the states $\vert r_{\pm}\rangle$ and $\vert s_{\pm}\rangle$ have to be discriminated at the two different paths of the setup, respectively. For this task HWP3, HWP3$'$ are set to turn the polarization of the light by an angle $\theta_{3}=\arcsin{(\tan{\alpha})}$ and HWP4, HWP4$'$ by $\theta_{4}=\frac{\pi}{2}$ \cite{Herzog2010}. The inconclusive results are detected at the avalanche photodiodes APD0 and APD0$'$. At the output of each interferometer the states $\vert r_+\rangle$, $\vert s_+\rangle$ or $\vert r_-\rangle$, $\vert s_-\rangle$ are transformed to orthogonal states collinear to $\vert V\rangle_{3}\pm\vert H\rangle_{3}$ and $\vert V\rangle_{4}\pm\vert H\rangle_{4}$ where + (-) applies for the case that $\rho_1$ ($\rho_2$) was prepared initially. By rotating the interferometer outcomes at HWP5, HWP5$'$ by angles of $\theta_{5}=\frac{\pi}{4}$, $\theta_{5'}=-\frac{\pi}{4}$ the conclusive results for input states $\rho_1$ ($\rho_2$) are detected at APD1 (APD2). The setup is equivalent to the one proposed in \cite{Herzog2010}, but requires less detectors. Due to the fact that in each interferometer the light in boths arms passes the same mirrors and beam splitters the interferometers are very stable. We could repeat the measurement on following days without any realignment.

\begin{figure}
\includegraphics[width=6cm]{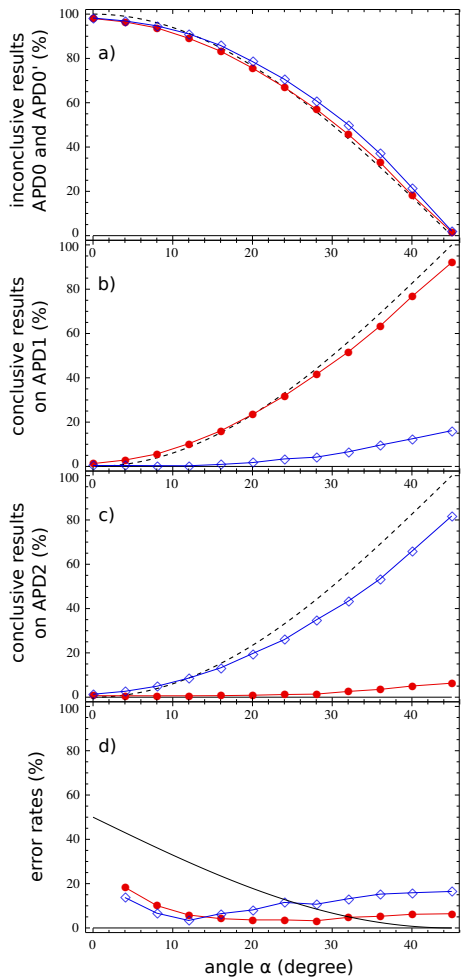}
\caption{Results for unambiguous state discrimination of two mixed single-photon states against the half angle of separation $\alpha$ for the cases that $\rho_{1}$ (circles) and $\rho_{2}$ (diamonds) were prepared. The dashed lines show the theoretical curves. (a) Percentage of inconclusive results, measured on APD0 and APD0$'$. (b), (c) Percentage of conclusive results, measured on APD1 and APD2, respectively. (d) Error rates for the conclusive results. The black line is the theoretical limit for the von Neumann measurement achieving minimum error discrimination (see text).}
\label{USD-results}
\end{figure}

In Figs. \ref{USD-results}(a)-(c) the measurement results are shown for the two mixed input states $\rho_1$ and $\rho_2$. All count rates are given in percentage of the sum of count rates on the four APDs. The total count rate was typically around 2000 $\mathrm{s}^{-1}$.  Also shown are the theoretical curves, given by $Q_{opt} = \cos 2\alpha$ for the relative number of inconclusive results and $1-Q_{opt}$ for the conclusive ones. The nonvanishing count rate on APD1 (APD2) for the case that $\rho_{2}$ ($\rho_{1}$) was prepared initially is mainly due to imperfections of the optical elements, especially the half-wave plates. The experimental error rates are the number of counts on the ``wrong`` APD divided by the number of all conclusive results, i.e. the counts on APD2 (APD1) divided by the counts on APD1 and APD2 for the case that $\rho_1$ ($\rho_2$) was prepared initially. These error rates are shown in Fig. \ref{USD-results}(d). Also shown is the theoretical error rate $P_E=\frac{1}{2}-\frac{1}{4}\mathrm{Tr}\vert \rho_{1}-\rho_{2}\vert$ for the von Neumann measurement which discriminates $\rho_{1}$ and $\rho_{2}$ with minimum error probability and without inconclusive results (black line). For angles $4^{\circ}<\alpha<32^{\circ}$ ($\rho_{1}$ prepared) and $4^{\circ}<\alpha<24^{\circ}$ ($\rho_{2}$ prepared) the error rates achieved here experimentally by following our unambiguous state discrimination strategy is smaller than the theoretical error rate for the projective measurement achieving minimum error would be. The difference in error rates for the two different input states arises from a slight misalignment of the beam when HWP2 and HWP2$'$ are inserted.

In conclusion, we demonstrated the applicability of our true single-photon source based on a single QD as an appropriate nonclassical light source for testing concepts of quantum state analysis. We performed the first experimental realization of optimum maximum-confidence discrimination and optimum unambiguous discrimination between two quantum states which both are mixed. These strategies for discriminating mixed states can be implemented in quantum information networks where decoherence leading to a mixing of pure states is inevitable.

\begin{acknowledgments}
Financial support by the DFG (SFB 787) is acknowledged. 
\end{acknowledgments}

\end{document}